\documentclass{appolb}
\usepackage{epsfig}
\usepackage{graphicx}

\def\sqeeb{\ifmmode{\sqrt{s_{\protect\bf\mathrm{ee}}}}\else
  {$\sqrt{s_{\protect\bf\mathrm{ee}}}$}\fi}
\def\epem{\ifmmode{\mathrm{e}^{+}\mathrm{e}^{-}}\else
  {$\mathrm{e}^{+}\mathrm{e}^{-}$}\fi}
\def\sqee{\ifmmode{\sqrt{s_\mathrm{ee}}}\else
  {$\sqrt{s_\mathrm{ee}}$}\fi}
\def\kp{{\ifmmode{k_{\perp}}\else{$k_{\perp}$}\fi}}
\def\etmean{\ifmmode{\bar{E}^{\mathrm{jet}}_{\mathrm{T}}}\else
  {$\bar{E}^{\mathrm{jet}}_{\mathrm{T}}$}\fi}
\def\etajmean{\ifmmode{|\bar{\eta}^\mathrm{jet}|}\else 
  {$|\bar{\eta}^\mathrm{jet}|$}\fi}
\def\lxg{\ifmmode{\mathrm{log_{10}}(x_{\gamma})}\else
  {${\mathrm{log_{10}}(x_{\gamma})}$}\fi}
\def\xg{\ifmmode{x_{\gamma}}\else{${x_{\gamma}}$}\fi}
\def\xgp{\ifmmode{x_{\gamma}^+}\else{${x_{\gamma}^+}$}\fi}
\def\xgm{\ifmmode{x_{\gamma}^-}\else{${x_{\gamma}^-}$}\fi}
\def\xgpm{\ifmmode{x_{\gamma}^{\pm}}\else 
  {${x_{\gamma}^{\pm}}$}\fi}
\def\ipb{\ifmmode {\mathrm{pb}^{-1}}\else 
  {$\mathrm{pb}^{-1}$}\fi}

\def\ee{\ifmmode{\mbox{e}^+\mbox{e}^-}\else
  {$\mbox{e}^+\mbox{e}^-$}\fi}
\def\thetamaxp{\ifmmode{\theta_\mathrm{max}'}\else
  {$\theta_\mathrm{max}'$}\fi}
\def\pt{\ifmmode{p_\mathrm{T}}\else{$p_\mathrm{T}$}\fi}
\def\Zzero{\ifmmode{\mathrm{Z}^{0}}\else{$\mathrm{Z}^{0}$}\fi}

\def\etjet{\ifmmode{E^\mathrm{jet}_\mathrm{T}}\else 
  {$E^\mathrm{jet}_\mathrm{T}$}\fi}
\def\ptmiss{\ifmmode{{P}_{\mathrm{T,MISS}}}\else 
  {${P}_{\mathrm{T,MISS}}$}\fi}
\def\mj1h2{\ifmmode{M_{\mathrm{J1H2}}}\else 
  {$M_{\mathrm{J1H2}}$}\fi}
\def\ebeam{\ifmmode{E_{\mathrm{BEAM}}}\else 
  {$E_{\mathrm{BEAM}}$}\fi}
\def\etajet{\ifmmode{|\eta^\mathrm{jet}|}\else 
  {$|\eta^\mathrm{jet}|$}\fi}
\def\etaj{\ifmmode{\eta^\mathrm{jet}}\else 
  {$\eta^\mathrm{jet}$}\fi}
\def\etajdef{\ifmmode{\eta^\mathrm{jet} = 
    -\ln\tan(\theta^\mathrm{jet}/2)}\else{$\eta^\mathrm{jet} = 
    -\ln\tan(\theta^\mathrm{jet}/2)$}\fi}
\def\detajet{\ifmmode{|\Delta\eta^\mathrm{jet}|}\else 
  {$|\Delta\eta^\mathrm{jet}|$}\fi}
\def\costhst{\ifmmode{|\mathrm{cos}\,\Theta^{*}|}\else 
  {$|\mathrm{cos}\,\Theta^{*}|$}\fi}
\def\etajetc{\ifmmode{|\eta^\mathrm{jet}_\mathrm{cntr}|}\else 
  {$|\eta^\mathrm{jet}_\mathrm{cntr}|$}\fi}
\def\etajetf{\ifmmode{|\eta^\mathrm{jet}_\mathrm{fwd}|}\else 
  {$|\eta^\mathrm{jet}_\mathrm{fwd}|$}\fi}
\def\etah{\ifmmode {\hat{\eta}}\else{$\hat{\eta}$}\fi}

\def\dsdeta{\ifmmode{\frac{\mathrm{d}{\sigma}_{\mathrm{dijet}}}
  {\mathrm{d}\etajet}}\else
    {$\frac{\mathrm{d}{\sigma}_{\mathrm{dijet}}}
      {\mathrm{d}\etajet}$}\fi}

\def\et{\ifmmode{E_\mathrm{T}}\else{$E_\mathrm{T}$}\fi}

\def\gg{\ifmmode{\gamma\gamma}\else{$\gamma\gamma$}\fi}
\def\gsg{\ifmmode{\gamma^{\star}\gamma}\else
  {$\gamma^{\star}\gamma$}\fi}
\def\PTMIA{\ifmmode{p_\mathrm{t}^\mathrm{mi}}\else
  {$p_\mathrm{t}^\mathrm{mi}$}\fi}
\def\sas1d{SaS\,1D}

\def\grvnlo{GRV\,HO}

\def\gs96nlo{GS96\,HO}
\def\lac1{LAC\,1}
\def\hadcor{\ifmmode{(1+\delta_{hadr})}\else{$(1+\delta_{hadr})$}\fi}

\begin{document}
\title{High momentum particle and jet production in photon-photon collisions
\thanks{Presented at the PHOTON2005 conference, September 2005, Warsaw, Poland}%
}
\author{Thorsten Wengler
\address{
University of Manchester\\
School of Physics and Astronomy\\
Manchester M13 9PL\\
United Kingdom}
}
\maketitle
\begin{abstract}

Jet and particle production have been studied in collisions of
quasi-real photons collected during the LEP2 program. OPAL and DELPHI
report good agreement of NLO perturbative QCD with the measured
differential di-jet cross sections, which reach a mean transverse
energy of the di-jet system of 25\,GeV. L3, on the other hand, finds
drastic disagreement of the same calculation with single jet
production for transverse jet momenta larger than about 25\,GeV. L3
observes similar disagreement between data and NLO QCD in their
measurements of charged and neutral particle production at high
transverse momenta of the particles. A recent measurement performed by
DELPHI of the same quantities does not confirm this observation.

\end{abstract}


\section{Introduction}

The large amount of hadronic photon-photon interactions at high
energies recorded by the LEP collaborations during the LEPII program
has led to several measurements that permit us to investigate the
validity of perturbative QCD for such processes. The emerging picture
confirms QCD as the correct theory to describe these
interactions. There are, however, several disagreements that still
need to be understood. This article will review the current state of
affairs and highlight areas where more work needs to be done. Many
more observables than can be discussed here are available in the
original publications. In all measurements discussed in this article
the photons entering the photon-photon collisions are quasi-real.

\section{Di-jet production}

DELPHI~\cite{bib-del-dijets} and OPAL~\cite{bib-opal-dijets} have
studied the production of di-jets at {\epem} centre-of-mass energies
{\sqee} from 189 to 209\,GeV, with a total integrated luminosity of
\mbox{550\,\ipb} and \mbox{593\,\ipb}, respectively.  Di-jet events
are of particular interest, as the two jets can be used to estimate
the fraction of the photon momentum participating in the hard
interaction, {\xg}, which is a sensitive probe of the structure of the
photon. The {\kp}-clustering algorithm\,{\cite{bib-ktclus}} is used
for the measurement of the differential cross-sections, because of the
advantages of this algorithm in comparing to theoretical
calculations\,{\cite{bib-ktisbest}}.

In leading order (LO) QCD, neglecting multiple parton interactions,
two hard parton jets are produced in {\gg} interactions.  In single-
or double-resolved interactions, these jets are expected to be
accompanied by one or two remnant jets.  A pair of variables, {\xgp}
and {\xgm}, can be defined that estimate the fraction of the photon's
momentum participating in the hard scattering:
\begin{equation}
\xgpm \equiv \frac{\displaystyle{\sum_{\rm jets=1,2}
 (E^\mathrm{jet}{\pm}p_z^\mathrm{jet})}}
 {{\displaystyle\sum_{\rm hfs}(E{\pm}p_z)}} ,
\label{eq-xgpm}
\end{equation}
where $p_z$ is the momentum component along the $z$ axis of the
detector and $E$ is the energy of the jets or objects of the
hadronic final state (hfs).  In LO, for direct events, all energy
of the event is contained in two jets, i.e.,~${\xgp}=1$ and
${\xgm}=1$, whereas for single-resolved or double-resolved events
one or both values are smaller than~1. Differential cross sections
as a function of {\xg} or in regions of {\xg} are therefore a
sensitive probe of the structure of the photon.

The experimental results are compared to a perturbative QCD
calculation at next-to-leading order (NLO)~\cite{bib-ggnlo} which uses
the {\grvnlo} parametrisation of the parton distribution functions of
the photon\,\cite{bib-grv}. OPAL applies the average of the
hadronisation corrections estimated by PYTHIA~\cite{bib-pythia} and
HERWIG~\cite{bib-herwig} to the calculation for this comparison.
DELPHI estimates the hadronisation corrections with PYTHIA and applies
them to the data.

\begin{figure}[thb]
\includegraphics[width=0.49\textwidth]{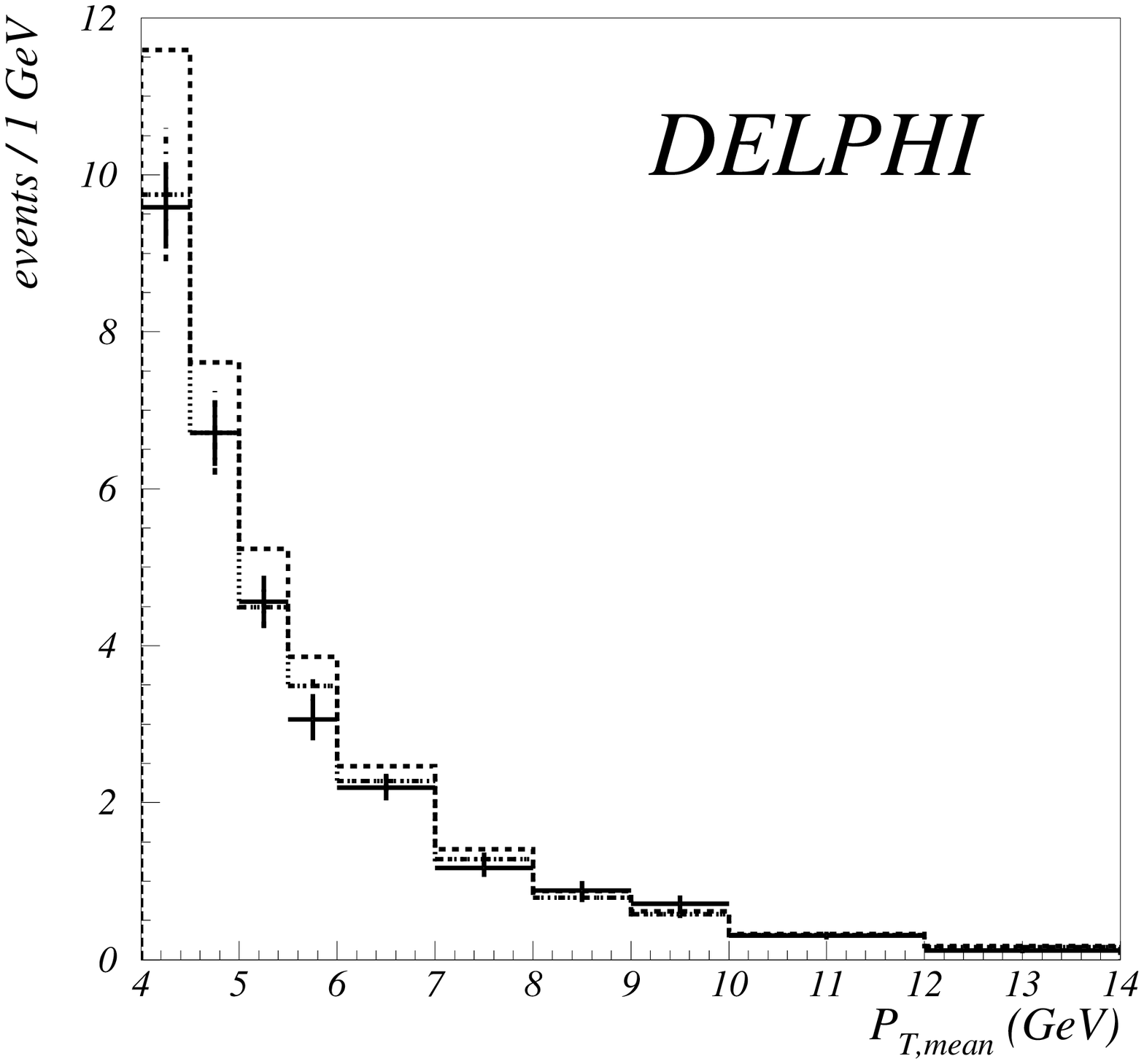}
\includegraphics[width=0.49\textwidth]{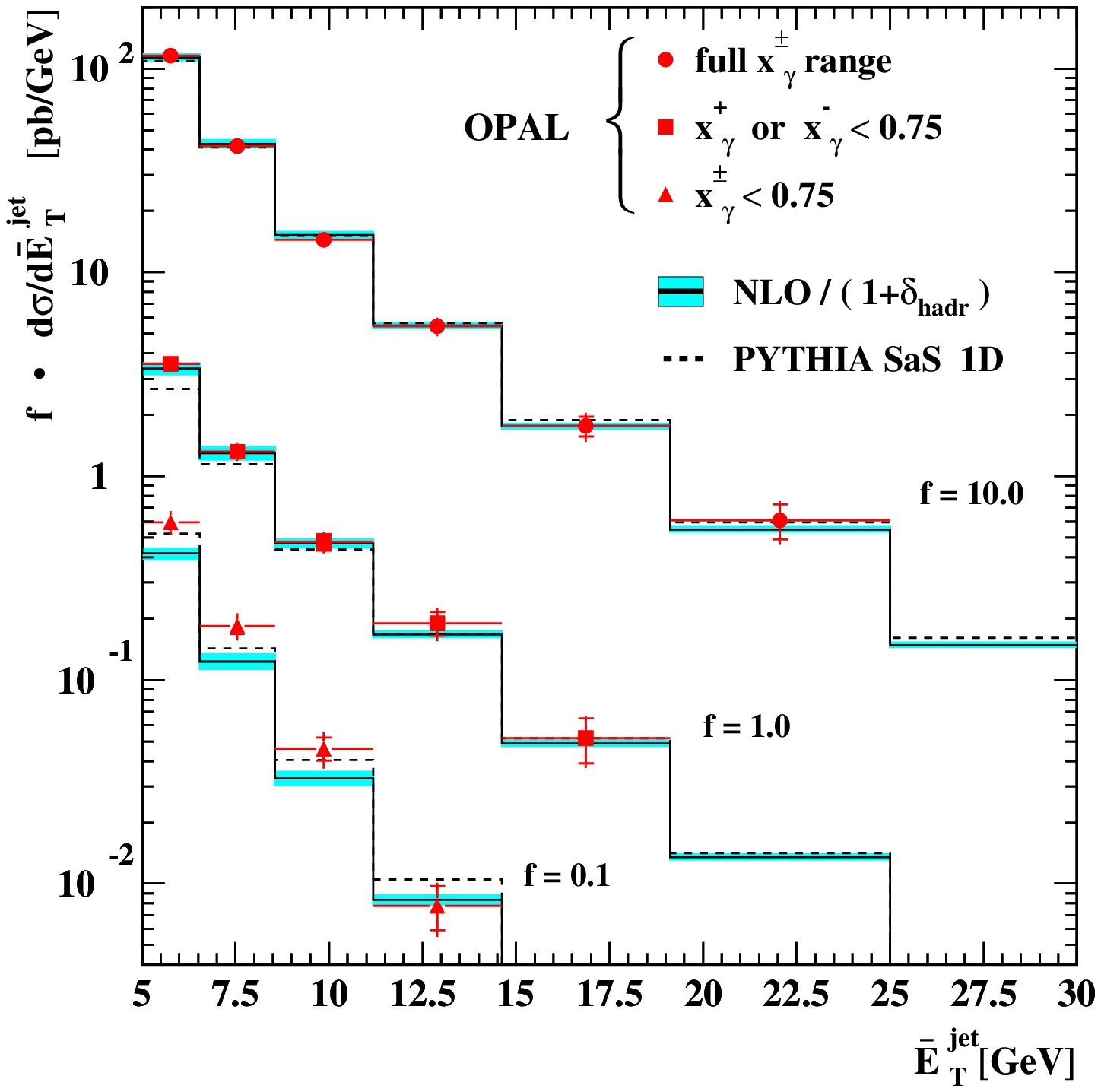}
\caption{Left: The di-jet cross section as a function of $P_{T,mean}$
as measured by DELPHI, compared to the LO (dashed) and NLO (dotted)
QCD calculation~\cite{bib-ggnlo}. Right: The di-jet cross-section as a
function of $\etmean$ as measuremd by OPAL, for the three regions in
{\xgp}-{\xgm}-space given in the figure.  The factor $f$ is used to
separate the three measurements in the figure more clearly.}
\label{fig:etmxs}
\end{figure}

The differential di-jet cross-section as a function of the mean
transverse momentum or energy of the di-jet system is shown in
Fig.\,{\ref{fig:etmxs}}. Both DELPHI and OPAL observe good agreement
with NLO QCD when integrating over {\xg}. The cross section predicted
for the case {\xgp} or {\xgm} $< 0.75$ is also in good agreement with
the OPAL measurement.  OPAL observes a significantly softer spectrum
for ${\xgpm} < 0.75$ than for the full {\xgp}-{\xgm}-space, as
expected from the dominance of resolved processes in this region. The
cross section predicted by NLO QCD is here below the measurement. It
should be noted that in this region the contribution from the
underlying event, not included in the calculation, is expected to be
largest, as shown below. PYTHIA 6.161 is in good agreement with the
OPAL data using the {\sas1d}~\cite{bib-sas} parton densities. PYTHIA
includes a model of the underlying event using multiple parton
interactions (MIA).

\begin{figure}[htb]
\includegraphics[width=0.52\textwidth]{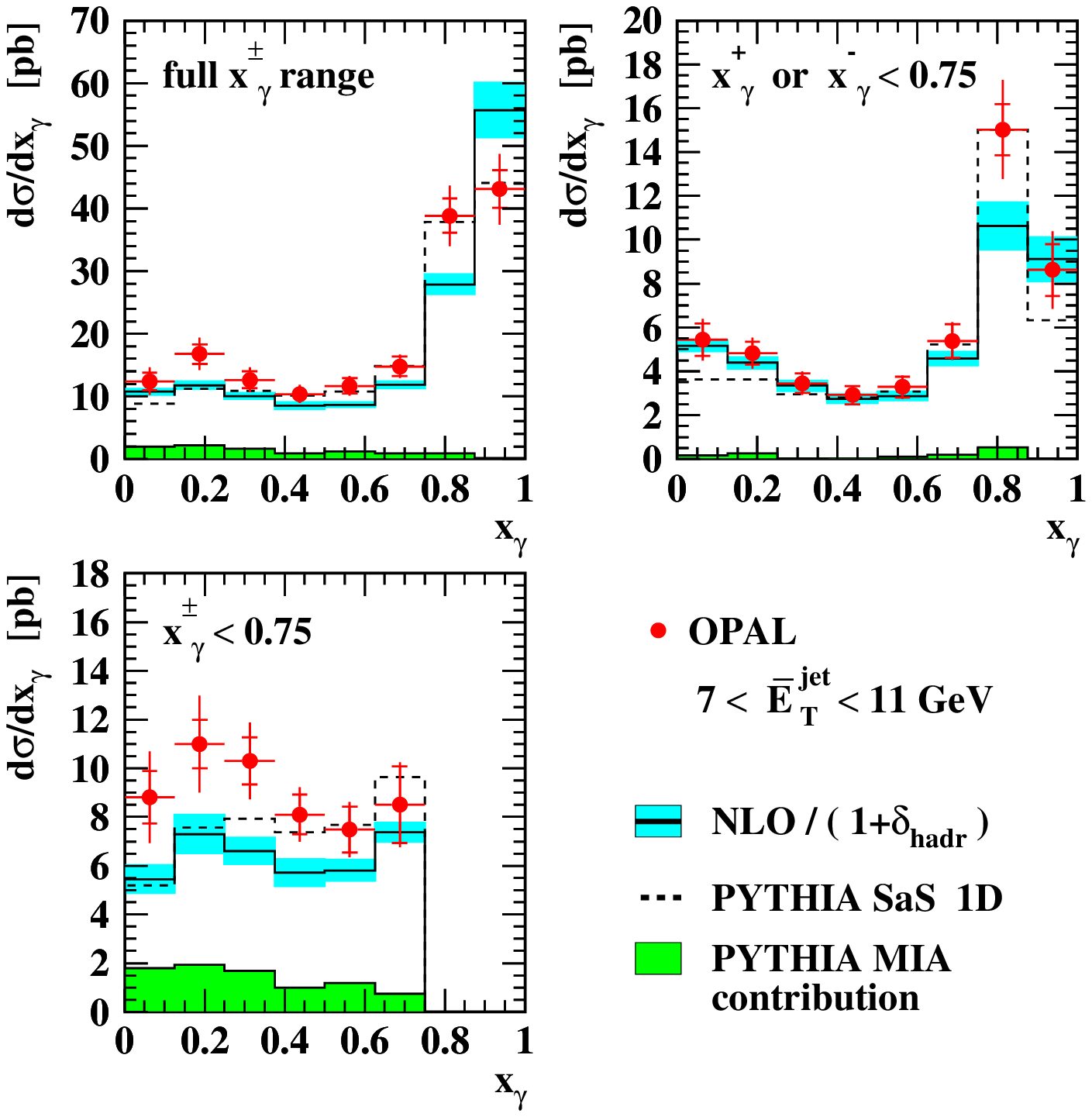}
\includegraphics[width=0.45\textwidth]{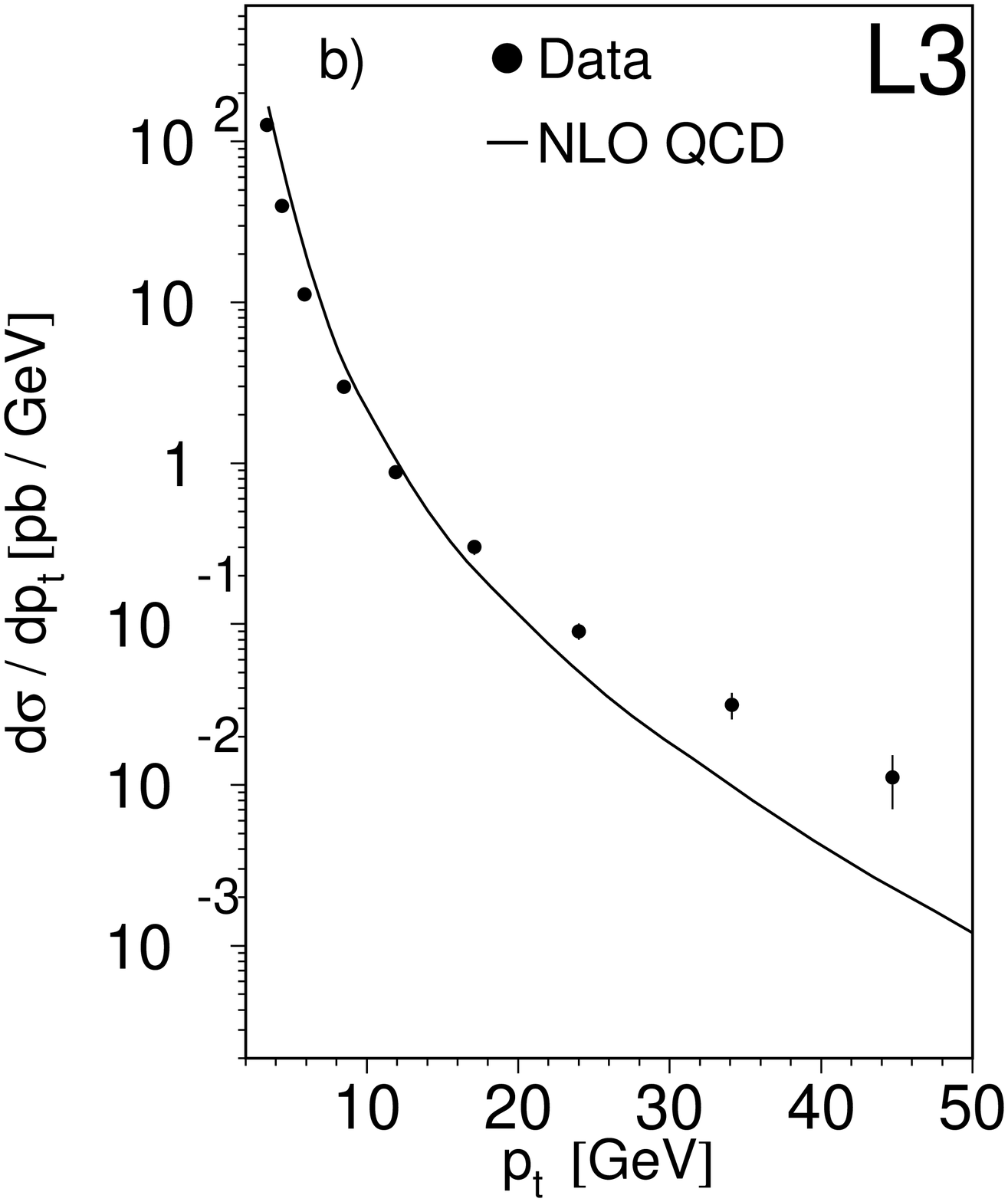}
\caption{Left: The di-jet cross-section measured by OPAL as a function
of $\xg$ and for the regions of the mean transverse energy $\etmean$
and {\xgpm} of the di-jet system indicated in the figures. Right: The
inclusive single jet cross section as measured by L3 compared to
NLO QCD. The theoretical scale uncertainty is less than 20\%.  }
\label{fig:xgcomb}
\end{figure}

The three plots on the left hand side of Fig.\,{\ref{fig:xgcomb}}
show the differential cross section measured by OPAL as a function of
{\xg} for the three regions in {\xgp}-{\xgm}-space described
above. The shaded histogram on the bottom of each of the three plots
indicates the contribution of MIA to the cross section as obtained
from the PYTHIA MC generator. It is evident especially for
${\xgpm}<0.75$ that the MIA contribution is of about the same size as
the discrepancy between the measurement and the NLO
prediction. Furthermore it is interesting to observe that there is
next to no MIA contribution to the cross section if either {\xgp} or
{\xgm} is required to be less than one, while the sensitivity to the
photon structure at small {\xg} is retained. As one would expect also
the agreement of the NLO calculation with the measurement is best in
this case. For large {\xg} the NLO calculation does not agree well
with the data. However, it has been pointed out that the calculation
of the cross section becomes increasingly problematic when approaching
${\xg=1}$\,\cite{bib-frixber,bib-klasenrev}.

Especially with these last measurements one is able to disentangle the
hard subprocess from soft contributions and make the firm statement
that NLO perturbative QCD is adequate to describe di-jet production in
photon-photon collisions in the regions of phase space where the
calculation can be expected to be complete and reliable, i.e. where
MIA contributions are small and for {\xg} not too close to unity. At
the same time a different sub-set of observables can be used to study
in more detail the nature of the soft processes leading to the
underlying event.

\section{Single jet inclusive production}

The L3 collaboration has measured inclusive jet production in
photon-photon interactions~\cite{bib-l3-jets}. A total integrated
luminosity of 560\,{\ipb} recorded at {\epem} centre-of-mass energies
$\sqee = 189-209$\,GeV is used.  Jets are reconstructed using the
{\kp}-clustering algorithm and analysed in the pseudorapidity range
$|\eta| < 1$ for jet transverse momenta $3 < p_t < 50$\,GeV. The
remaining background from other processes after event selection
increases from about 5\% at low $p_t$ to about 20\% at high
$p_t$. This background is subtracted bin-by-bin from the data before
corrections for selection efficiency and detector acceptance are
applied. The differential cross section as a function of $p_t$ is
shown in Fig.\,\ref{fig:xgcomb}. The distribution can be described
by a power law function $Ap_{t}^{-B}$ with $B=3.6\pm0.1$. A comparison
to an NLO perturbative QCD calculation~\cite{bib-frixber} using the
{\grvnlo} parton density functions fails to describe the data for jet
transverse momenta larger than about 25\,GeV.  This calculation was
shown to be in agreement with the calculation compared to the di-jet
observables above. There is clearly a need here for additional data to
be published to confirm or contradict the observed discrepancy.

\section{Hadron production}

\begin{figure}
\label{sngl}
\includegraphics[width=0.97\textwidth]{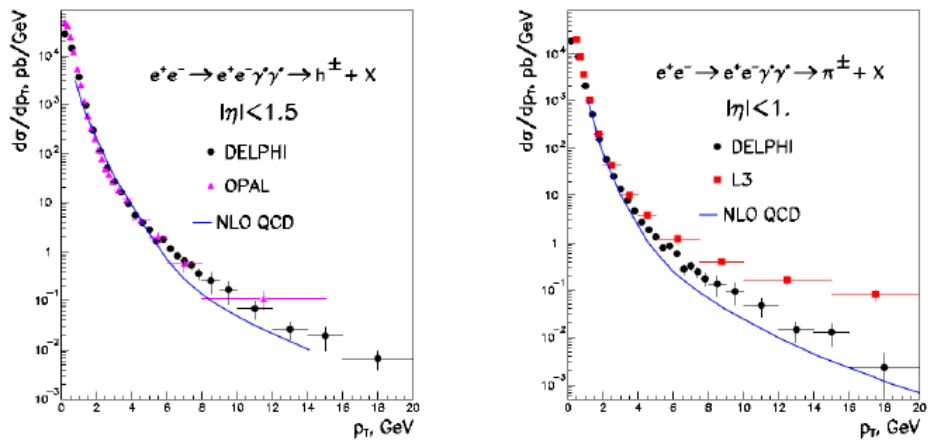}
\caption{Transverse momentum spectra of charged hadrons as measured by
DELPHI and OPAL (left) and of charged pions as measured by DELPHI and
L3 (right). In both cases the measurement are compared to a
calculation in NLO QCD~\cite{bib-ggnlo-had}. Figure taken
from~\cite{bib-del-ichep04}.}
\label{fig:had}      
\end{figure}

L3 has studied the production of charged and neutral pions in
414\,{\ipb} of data recorded at $e^+e^-$ centre-of-mass energies
$\sqee = 189-209$\,GeV. The right plot in Fig.~\ref{fig:had} 
shows the transverse momentum spectrum of charged pions. It is
evident that the corresponding calculation in NLO QCD fails to
describe the data for momenta larger than about 4\,GeV. At the
highest charged particle momenta measured the theory underestimates
the data by more than an order of magnitude. A similar measurement
by L3 of neutral pion production~\cite{bib-l3-pizero} leads to the
same conclusions. Yet the presence of high momentum particles should
indicate the presence of a hard scale, and the perturbative
calculation should be reliable. The discrepancy can therefore not
easily be understood in terms of the NLO calculation on parton
level. Furthermore at high momenta the interaction of two photons is
expected to be dominated by the so-called direct process, i.e. the
exchange of a fermion, such that uncertainties in the knowledge of
the photon structure are not expected to be very important. The
experimental background at the largest momenta measured has been
estimated from MC to be about 20\%, and is subtracted from the data
before the comparisons with theory are carried out. To compare the
parton level calculation to the data, it is folded with the
appropriate fragmentation function.  It is not expected that
fragmentation effects could explain discrepancies of this magnitude.
Furthermore in the study of single jet production described above,
which is sensitive to the same partonic processes, no fragmentation
functions are used in comparing data and theory. Hadronisation
corrections in the upper half of the transverse momentum spectrum
shown in the right plot of Fig.~\ref{fig:xgcomb} are estimated to be
below 10\%.  One has to conclude that these discrepancies seen in
the L3 data are very significant, and at present not understood.

DELPHI has performed a similar measurement of charged hadron
production in 617\,{\ipb} of data with centre-of-mass energies $\sqee
= 161-209$\,GeV~\cite{bib-del-had}.  The right plot in
Fig.~\ref{fig:had} shows the DELPHI data in comparison with L3 and
the prediction of NLO QCD. The DELPHI measurement is clearly
incompatible with the L3 data and in much better agreement with NLO
QCD, if still somewhat above the calculation at larger transverse
momenta. In studying the background especially at the largest
available transverse momenta DELPHI finds that a more restrictive event
selection is needed in their case to reduce the background to an
acceptable level. This concerns in particular the upper limits on the
total energy or on the total invariant hadronic mass in the event used
to suppress background from hadronic decays of the $Z^{0}$. When
modelling their selection on the one used by L3, DELPHI observes the
high transverse momentum region to be dominated by this
background. While this may serve as an indication of the difficulties
involved in measuring the hadron spectra in the region of large
transverse momenta, it should be noted that the level of background to
a signal process remaining after a set of selection criteria have been
applied depends to some extend on the experimental apparatus, and
hence cannot be directly compared across experiments.

The left plot in Fig.~\ref{fig:had} compares the DELPHI data to an
older measurement of OPAL~\cite{bib-opal-had} at lower energies. The
OPAL data do not yet reach high enough in transverse momentum to
contribute to the discussion of the data-theory discrepancies observed
by L3. Measurements of high momentum hadron production from
ALEPH and OPAL using the full LEPII data set would clearly be very
interesting to see.

\end{document}